%% file: OJ287_VLBI_v11.0.tex
\documentclass[apj]{emulateapj}

\usepackage{natbib}
\usepackage{graphicx}
\usepackage{apjfonts}

\slugcomment{Accepted for publication in The Astrophysical Journal}

\shorttitle{Erratic Jet Wobbling in the BL~Lac Object \object{OJ287}}
\shortauthors{Agudo et al.}

\begin{document}

\title{Erratic Jet Wobbling in the BL~Lacertae Object \object{OJ287} Revealed by Sixteen Years of 7\,mm VLBA Observations}

\author{Iv\'{a}n Agudo\altaffilmark{1,2}, 
            Alan P.~Marscher\altaffilmark{2}, 
            Svetlana G.~Jorstad\altaffilmark{2,3}, 
            Jos\'{e} L.~G\'omez\altaffilmark{1}, 
            Manel Perucho\altaffilmark{4}, 
            B. Glenn Piner\altaffilmark{5},
            Mar\'ia Rioja\altaffilmark{6,7},
            Richard Dodson\altaffilmark{7}
            }


\altaffiltext{1}{Instituto de Astrof\'isica de Andaluc\'ia, CSIC, Apartado 3004, 18080, Granada, Spain; \email{iagudo@iaa.es}}

\altaffiltext{2}{Institute for Astrophysical Research, Boston University, 725 Commonwealth Avenue, Boston, MA 02215, USA}

\altaffiltext{3}{Astronomical Institute, St. Petersburg State University, Universitetskij Pr. 28, Petrodvorets, 198504 St. Petersburg, Russia}

\altaffiltext{4}{Departament d'Astronomia i Astrof\'isica, Universitat de Val\`encia, Dr. Moliner 50, E-46100	Burjassot,	Val\`encia, Spain}

\altaffiltext{5}{Department of Physics and Astronomy, Whittier College, 13406 E. Philadelphia Street, Whittier, CA 90608, USA}

\altaffiltext{6}{Observatorio Astron\'omico Nacional, Apdo. 112, E-28803 Alcal\'a de Henares, Madrid, Spain}

\altaffiltext{7}{ICRAR/University of Western Australia, 35 Stirling Highway, Crawley WA 6009, Australia}

\begin{abstract}
We present the results from an ultra-high-resolution 7\,mm Very Long Baseline Array (VLBA) study of the relativistic jet in the BL Lacertae object \object{OJ287} from 1995 to 2011 containing 136 total intensity images.
Analysis of the image sequence reveals a sharp jet-position-angle swing by $>100^{\circ}$ during [2004,2006], as viewed in the plane of the sky, that we interpret as the crossing of the jet from one side of the line of sight to the other during a softer and longer term swing of the inner jet.
Modulating such long term swing, our images also show for the first time a prominent erratic wobbling behavior of the innermost $\sim0.4$\,mas of the jet with fluctuations in position angle of up to $\sim40^{\circ}$ over time scales $\sim2$\,yr. 
This is accompanied by highly superluminal motions along non-radial trajectories, which reflect the remarkable non-ballistic nature of the jet plasma on these scales.
The erratic nature and short time scales of the observed behavior rules out scenarios such as binary black hole systems, accretion disk precession, and interaction with the ambient medium as possible origins of the phenomenon on the scales probed by our observations, although such processes may cause longer-term modulation of the jet direction.
We propose that variable asymmetric injection of the jet flow; perhaps related to turbulence in the accretion disk; coupled with hydrodynamic instabilities, leads to the non-ballistic dynamics that cause the observed non-periodic changes in the direction of the inner jet. 
\end{abstract}


\section{Introduction}
\label{Intr}
The most luminous long-lived sources of radiation in the cosmos---active galactic nuclei (AGN)---are powered by gas falling from an accretion disk onto a super-massive black hole (BH, $\gtrsim10^7\,\rm{M}_{\sun}$) at their center. 
AGN classified as blazars are characterized by wild variability of flux of non-thermal radiation from radio to $\gamma$-ray frequencies. 
Members of this class include BL~Lacertae objects (BL~Lacs) and, at higher emitted power, flat spectrum radio quasars (FSRQ). 
The remarkable properties of blazars include superluminal apparent motions that can exceed $40\,c$ \citep{Jorstad:2005p264}, as well as substantial changes in flux and linear polarization on time scales as short as minutes. 
These phenomena are thought to be caused by relativistic jets of highly energized, magnetized plasma that are propelled along the rotational poles of the BH-disk system \citep[e.g.,][]{Blandford:1977p6285}. 
AGN jets pointing within $\lesssim10^{\circ}$ of our line of sight  beam their radiation and shorten the variability time scales to give blazars their extreme properties.

High-resolution long-term very-long baseline interferometric (VLBI) monitoring of blazars is revealing a growing number of cases where the observed (in projection) jets undergo rapid swings of position angle in their innermost regions while remaining collimated out to kiloparsec scales or farther \citep[e.g.,][]{2003ApJ...589L...9H,Stirling:2003p154,Tateyama:2004p11950,Bach:2005p462,Lobanov:2005p15363,Mutel:2005p443,Savolainen:2006p13808,Agudo:2007p132}.
Time scales ranging between $\sim2$ and $\gtrsim20$ years and projected oscillations of the jet position angle with amplitudes from $\sim25^{\circ}$ to $\gtrsim100^{\circ}$ are typical for the reported cases \citep[see][for examples]{Agudo:2009p14514}.
This phenomenon---which has previously been termed \emph{jet precession}, although we prefer \emph{wobbling} given the doubts about its periodicity in some sources \citep[e.g.,][]{Mutel:2005p443}---has been reported so far only in the innermost (parsec-scale) regions of jets.
This already implies that the origin of jet wobbling might be a mechanism intimately tied to the origin of the jet.
Bends and helical structures in AGN are also claimed to be produced by changes in direction near the base of the jet \citep[e.g.][]{Dhawan:1998p15371,2003ApJ...584..135L,Hardee:2005p260,Agudo:2006p331}, which may be related to the wobbling phenomenon as well.

During the past few decades, the BL~Lac object \object{OJ287} ($z=0.306$) has shown quasi-periodic double-peaked optical flares every $\sim12$\,yr, most recently in $[2005,2008]$ \citep[e.g.,][]{Villforth:2010p11557}.  
This quasi-periodicity has been used to support binary BH models, as well as other scenarios such as accretion disk or jet instabilities \citep[see][for a review of these different scenarios]{Villforth:2010p11557}.
Consistent with the binary BH hypothesis, \citet{Tateyama:2004p11950} reported jet wobbling in a 3.5\,cm long-term Very Long Baseline Array (VLBA) study from 1994 to 2002. 
They interpreted this as the result of ballistic precession of the jet, also with a periodicity of $\sim12$\,yr.
Based on the \cite{Valtonen:2008p5346} binary BH model for \object{OJ287}, \citet{Valtonen:2011p15373} have proposed a new model for wobbling of the jet in this source with a periodicity of $\sim120$\,yr and a modulation of the jet position angle on a $\sim12$\,yr time scale driven by changes in orientation of the inner accretion disk of the proposed primary black hole.

To re-evaluate the origin of jet wobbling in general, and in \object{OJ287} in particular, we present here the results of a detailed 7\,mm VLBA study of \object{OJ287} from 1995 to 2011 that includes 136 VLBI images.
Owing to the reduced effect of synchrotron self-absorption in blazar jets at these short wavelengths, compared with centimeter-wave radio bands, and the ultra-high ($\sim0.15$\,milliarcsecond) angular resolution of the images, we scrutinize the inner jet regions in much finer detail than is possible at centimeter wavelengths.
We adopt hereafter the standard $\Lambda$CDM cosmology with $H_0$=71 km s$^{-1}$ Mpc$^{-1}$, $\Omega_M=0.27$, and $\Omega_\Lambda=0.73$.
Under this assumption, at the redshift of \object{OJ287}, 1\,milliarcsecond (mas) translates to a projected distance of $4.48$\,pc, and a proper motion of 1\,mas\,yr$^{-1}$ corresponds to a superluminal speed of $19.09\,c$.

\section{Observations, Data Reduction, and Modeling}

We present 136 total-intensity VLBA images of \object{OJ287} at 7\,mm obtained from January 1995 to June 2011; see Table~\ref{T1} and Fig.~\ref{maps}.  
Linear polarization 7\,mm VLBA images from December 2010 to June 2011 are also shown in Fig.~\ref{polmaps} \citep[see][for polarization maps before Dec. 2010]{Agudo:2011p14707}. 
Within the time range 1997.86-2001.95, most of the data were obtained on a monthly basis under a program that monitored the compact structure of the radio galaxy \object{3C~120} \citep{2000Sci...289.2317G,Gomez:2001p201,Gomez:2008p30,Gomez:2011p15108}, where \object{OJ287} was one of the main calibrators.
During the time frame 2006.38-2011.45, most of the observations were performed on a roughly monthly basis under the Boston University 7\,mm blazar monitoring program \citep{Darcangelo:2009p6953,Larionov:2008p338,Marscher:2010p11374,Jorstad:2010p11830,Agudo:2011p14707,Chatterjee:2011p15967,Agudo:2011p15946}.
The data from these programs are also supplemented by \object{OJ287} observations from several other independent 7\,mm VLBA programs dating back to 1995 \citep[e.g,][]{Jorstad:2001p5655,Agudo:2006p331,Piner:2006p12065,Agudo:2007p132,Agudo:2009p14514}. 
To cover gaps in the time coverage of the source, we obtained data at eight observing epochs from the VLBA archive. 

The data were reduced following the standard method for high frequency VLBA polarimetric data described in, e.g., \citet{Jorstad:2005p264,Agudo:2006p331}, and \citet{Gomez:2011p15108}.
For further details on the data reduction procedure followed for each particular data set, we refer to the corresponding publications indicated in Table~\ref{T1}.

To make a simpler representation of the brightness distribution in our images, we have performed model fits in which the emission features are represented by circular Gaussian emission components, with the number of components kept as small as possible.
To estimate the uncertainties of the model fit parameters, we follow the prescription of \cite{Jorstad:2005p264}.
Our fits typically include no more than two or three compact components within the innermost $\sim0.2$-$0.4$\,mas of the jet (Table~\ref{T2}).
We use the position of these components in the inner $\sim0.4$\,mas of the observed jet to define the jet position angle (JPA) in  \object{OJ287} as observed from Earth (see Fig.~\ref{JPA}). 

\begin{figure*}
   \centering
   \includegraphics[clip,width=17.cm]{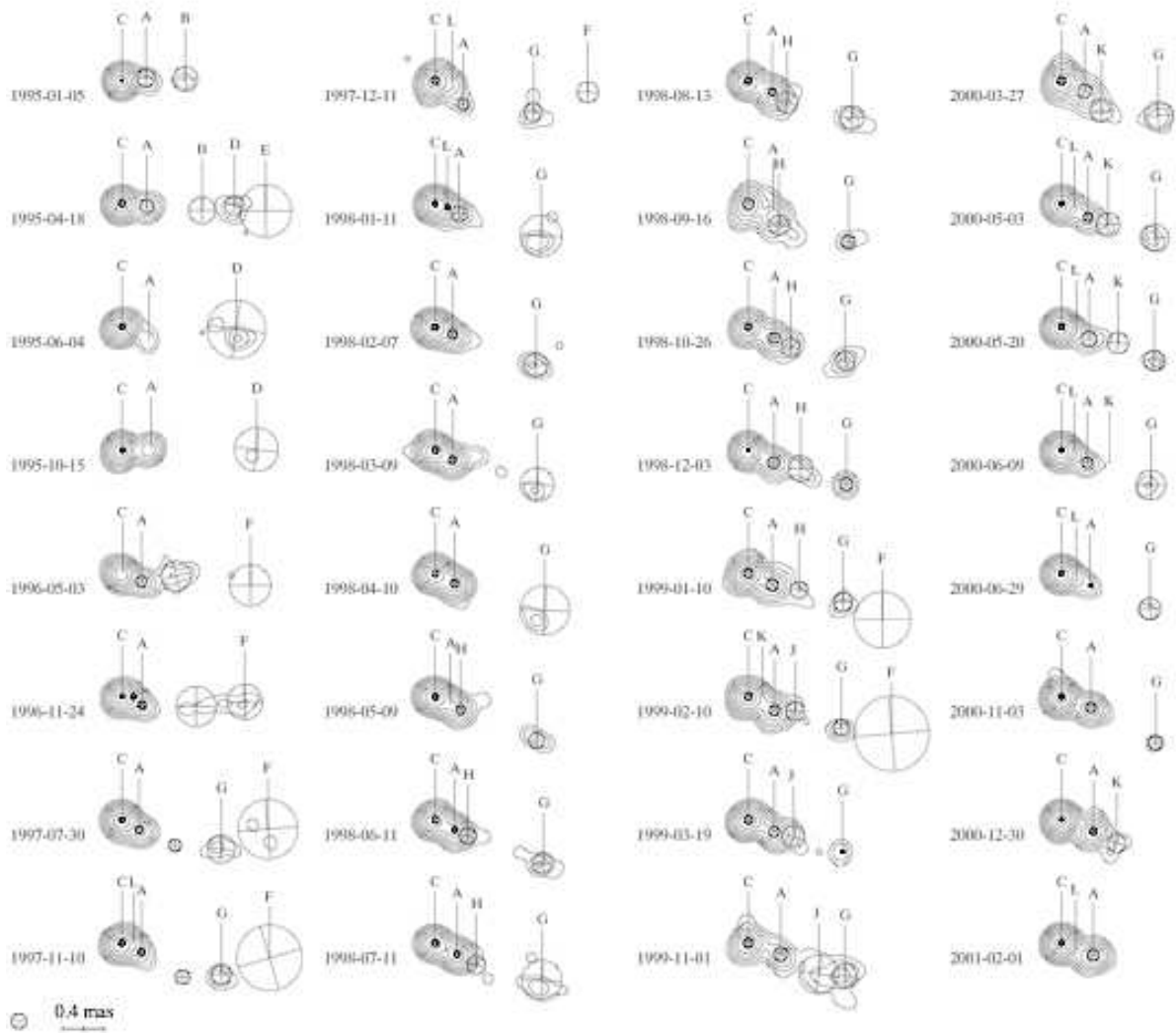}
   \caption{Sequence of 7\,mm VLBA maps of OJ287 from 1995 to 2011, convolved with a circular Gaussian beam with $\rm{FWHM}=0.15$\,mas (represented by the circle in the left-bottom corner). Contours symbolize the observed total intensity with levels in factors of 2 from $0.1$ to  $51.2$\,\% plus $90.0$\,\% of peak$=9.018$\,Jy/beam. A movie corresponding to this sequence of images can be found at URL {\tt https://w3.iaa.es/$\sim$iagudo/research/OJ287/OJ287LT.mov}.}
   \label{maps}
\end{figure*}

\addtocounter{figure}{-1}

\begin{figure*}
   \centering
   \includegraphics[clip,width=17.cm]{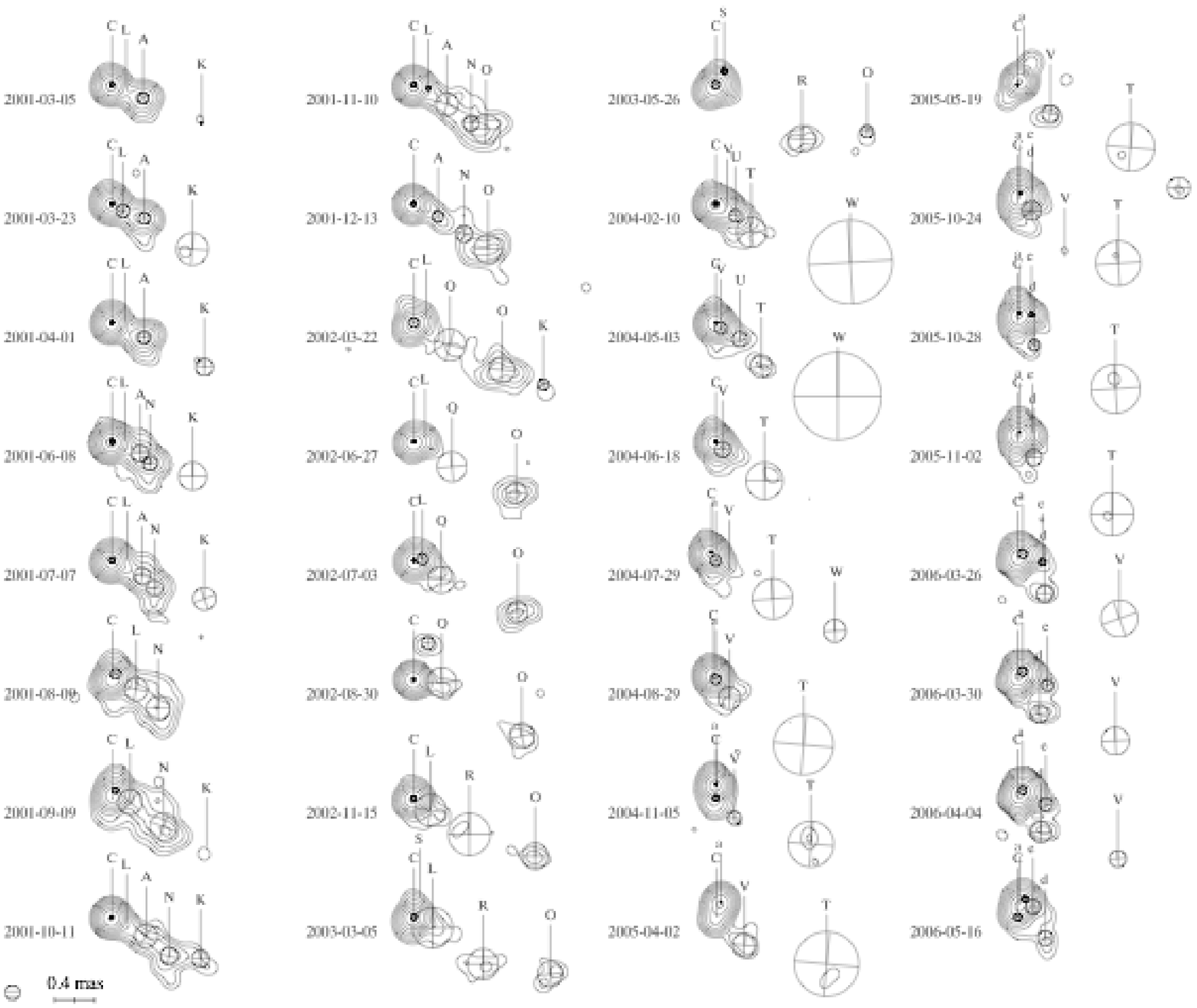}
   \caption{Continued.}
\end{figure*}

\addtocounter{figure}{-1}

\begin{figure*}
   \centering
   \includegraphics[clip,width=17.cm]{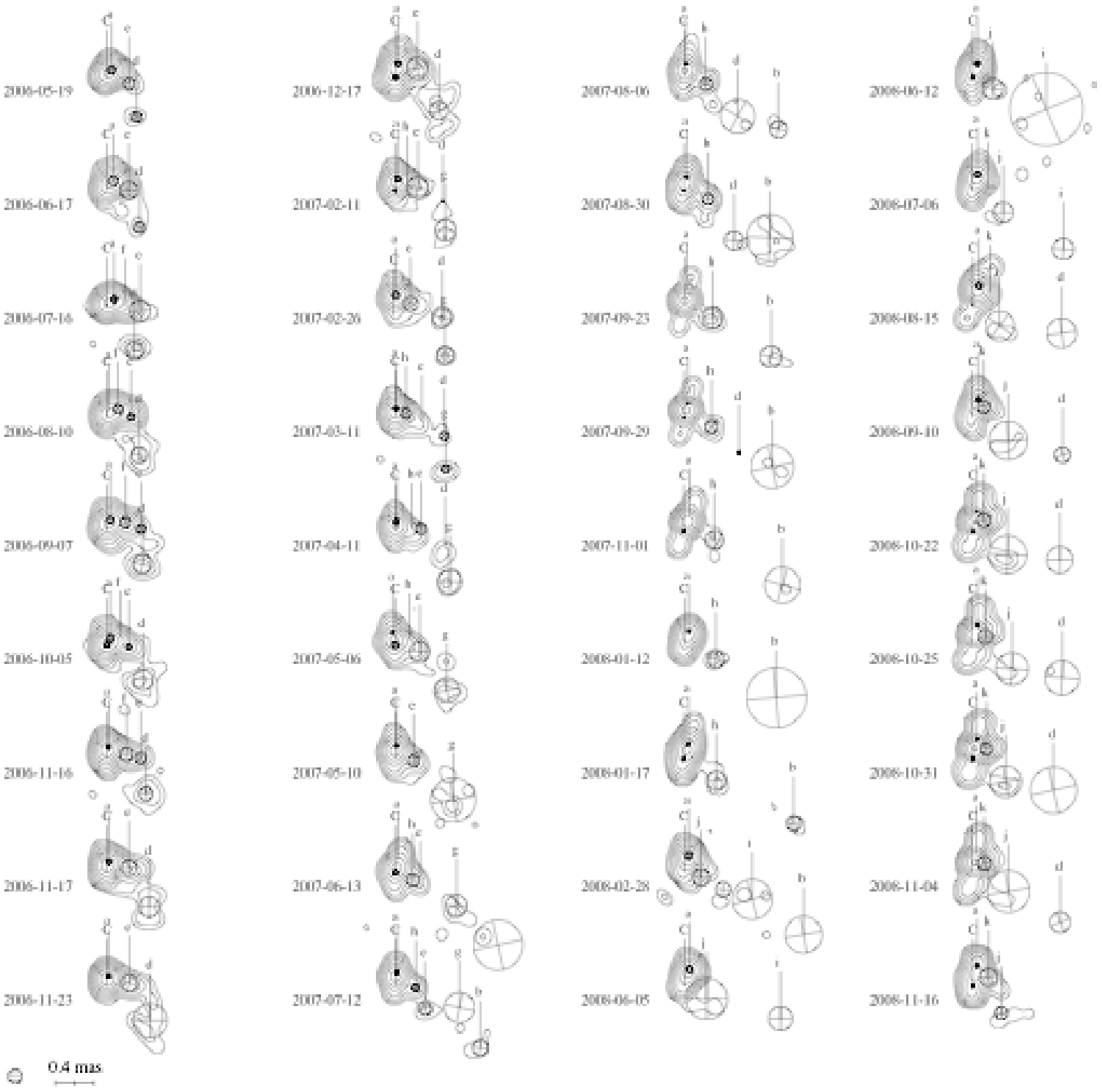}
   \caption{Continued.}
\end{figure*}

\addtocounter{figure}{-1}

\begin{figure*}
   \centering
   \includegraphics[clip,width=17.cm]{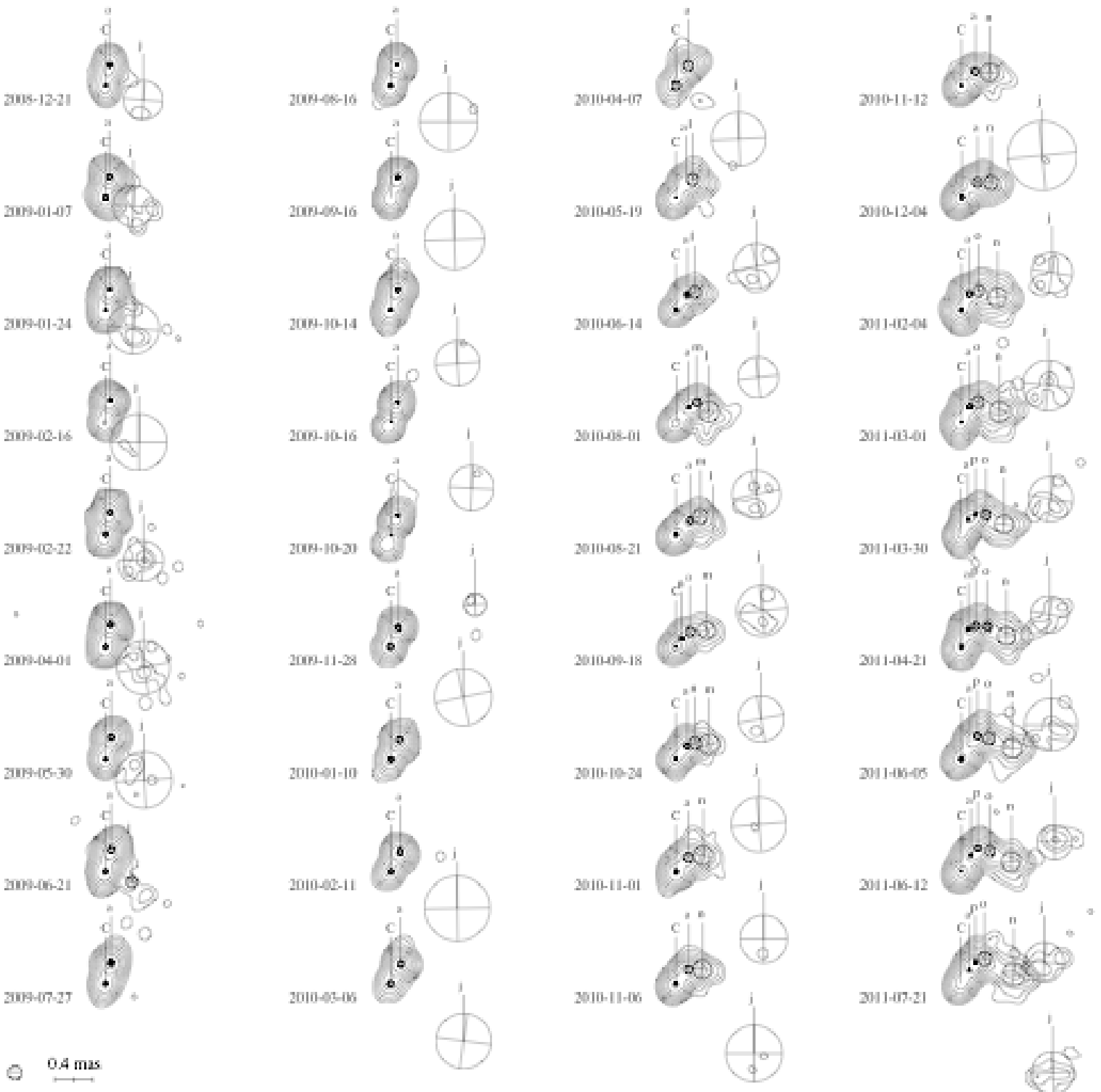}
   \caption{Continued.}
\end{figure*}


\section{Results and Analysis}

\subsection{The Innermost VLBI Jet Feature}

We define the ``core'' of the brightness distribution of \object{OJ287} as the innermost jet feature visible in our 7\,mm VLBA images.
Given that the core is not usually associated with moving blobs \citep[e.g.][]{Jorstad:2005p264} we assume that it is a stationary feature in the jet of \object{OJ287}.
Identification of the core in a multi-epoch kinematic VLBI study like ours---where self-calibrated VLBI images do not preserve the absolute positional information---is important to define a kinematic reference point of the system.

For the time period from the beginning of 1995 to mid-2004, the identification of the core is clear.
The bright jet feature at the eastern end (northeastern end at some epochs) of the brightness distribution, i.e., model-fit component {\it C} (see Fig.~\ref{maps}), from which superluminal features emerge and travel downstream (see Section~\ref{kin}), corresponds to the core.
However, the identification of the core is more complicated during the time range from mid-2004 to mid-2011.
Whereas the outer knots \citep[beyond $0.5$\,mas from component {\it  C}, labelled {\it  C0} in ][]{Agudo:2011p14707} continue propagating toward the southwest, the region within the innermost $\sim0.4$\,mas behaves in a completely different way than observed before. 
The motion of knots emerging from the core suddenly shifted to the north (and later to the northwest) direction after mid-2004. 
The new direction is nearly perpendicular to the typical jet position angle during the 1995-2003 time range (see Fig.~\ref{maps}).

\begin{figure}
   \centering
   \includegraphics[clip,width=5.cm]{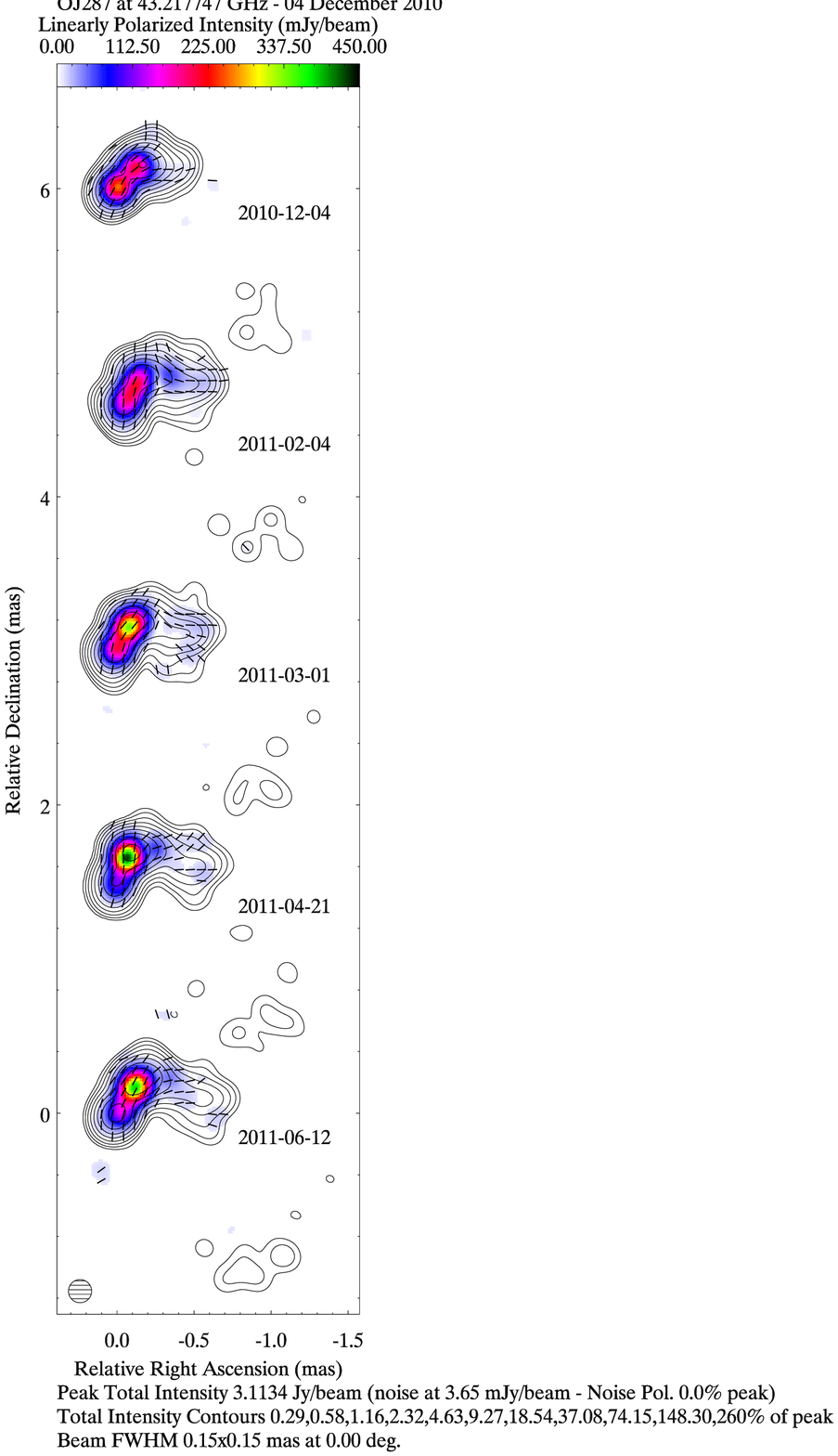}
   \caption{Similar to Fig.~\ref{maps} but restricted to the period December 2010 to June 2011. In addition, linearly polarized intensity (color scale) and polarization electric vector position angle distributions (short black sticks) are superimposed at each observing epoch.}
   \label{polmaps}
\end{figure}

The identification of the core with component {\it  C} is justified by:
(i) the detection of superluminal motion away from component {\it  C}, toward jet feature {\it  a}, and then outward (see component {\it  n})\footnote{this is also clearly seen in the evolution of the total intensity of the jet in an animation at URL {\tt https://w3.iaa.es/ $\sim$iagudo/research/OJ287/OJ287\_2010\_ejection.mov}, as well as in Figs.~\ref{maps} and \ref{polmaps} and Table~\ref{T3}.}; 
(ii) the progressive decrease of jet intensity toward the west of component {\it  a} (labelled {\it  C1} in \citet{Agudo:2011p14707}, see Fig.~\ref{maps}) at epochs after mid-2010;  
(iii) the confirmation of a continuous jet flow tracing the arc {\it  C-a-n} after mid-2010 in both total and polarized intensity and in polarization angle (which remains parallel to the local jet axis; see Fig.~\ref{polmaps});
and (iv) the smaller mean size and higher mean observed brightness-temperature\footnote{The observed brightness temperature was computed as in \citet{Marscher:1979p1340} and \citet{Agudo:2006p203}} of component {\it  C} from fall 2004 ($\phi(\rm{C})=(0.02\pm0.02)$\,mas and $T_{\rm{B}}(\rm{C})\sim2.7\times10^{12}$\,K, respectively.) as compared to that measured for component {\it  a} ($\phi(\rm{a})=(0.05\pm0.03)$\,mas and $T_{\rm{B}}(\rm{a})\sim1.1\times10^{12}$\,K) on the same dates.
All these lines of evidence together suggest that the jet flows downstream from component {\it  C} toward component {\it  a}, and at epochs after mid-2010 toward components {\it  m} and {\it  n}. 
This unambiguously identifies component {\it  C} with the innermost visible jet feature in our 7\,mm images, i.e., with the ``core'' of \object{OJ287} as defined above.

\subsection{Jet Wobbling in OJ287}
\label{OJ287wobbling}

\begin{figure}
   \centering
   \includegraphics[clip,width=8.5cm]{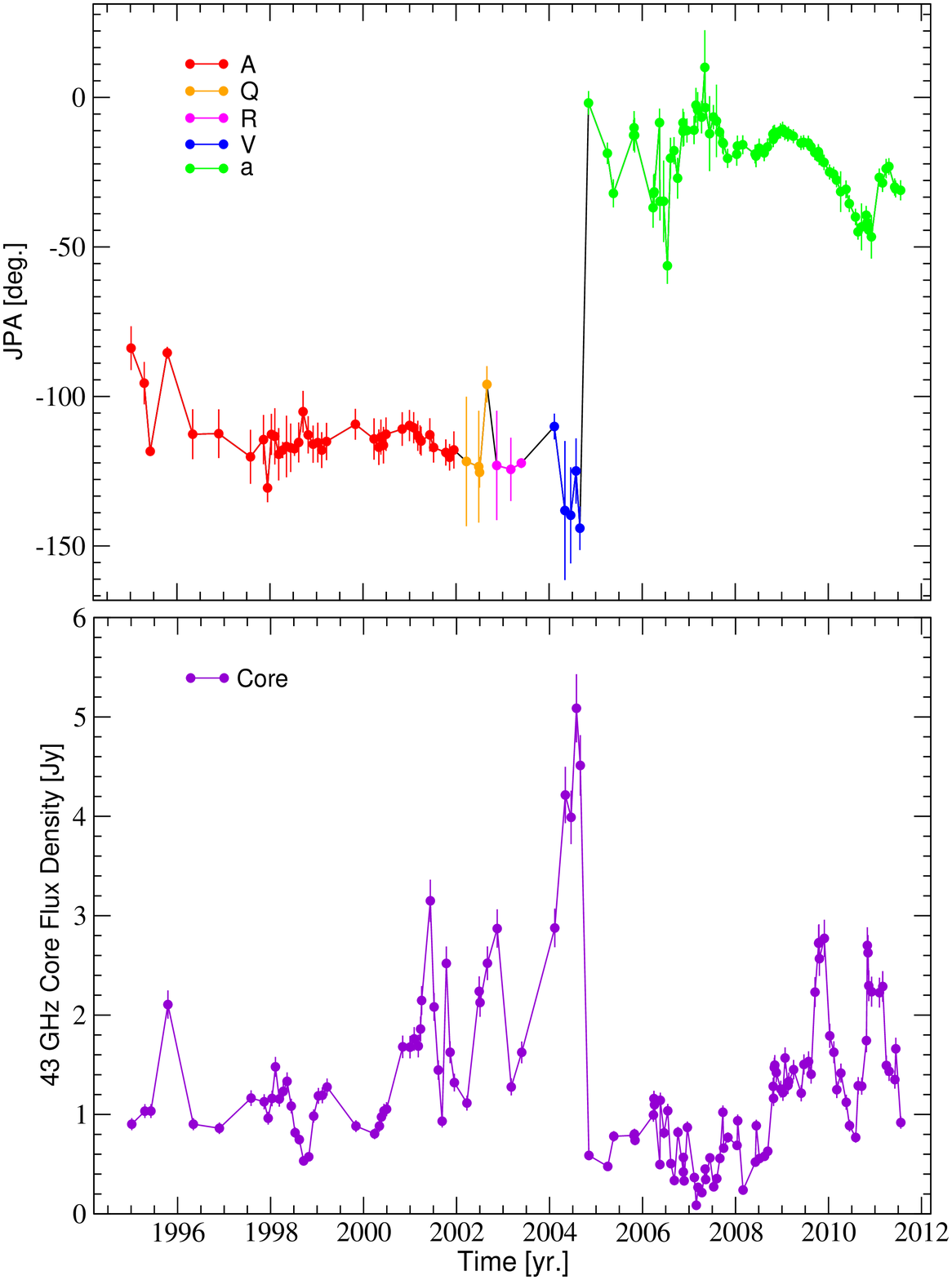}
   \caption{{\it Top:} Time evolution of the 7\,mm jet position angle (JPA) of OJ287. Different colors indicate the various jet features whose positions are used to estimate the JPA; see also Fig~\ref{maps}. {\it Bottom:} 7\,mm total flux density evolution of the core.}
   \label{JPA}
\end{figure}

Figures~\ref{maps} and \ref{JPA} show clear clockwise swing of the jet in \object{OJ287} over the time range between the beginning of 1995 (with JPA$\sim-100^{\circ}$) and mid-2004 (with JPA$\sim-150^{\circ}$).
This trend is modulated by shorter term jet wobbling on time scales $\sim2$\,yr (as measured from local JPA maxima and minima) and relatively small amplitude, $\gtrsim25^{\circ}$.
During the [1995,2004] time range, this phenomenology is not necessarily inconsistent with the interpretation of strictly periodic jet precession by \citet{Tateyama:2004p11950}, as long as the short-term wobbling on fine angular scales seen in our data is superposed on long-term, sinusoidal modulation (with period $\sim12$\,yr) as inferred by these authors on coarser angular scales. 
In addition to the wobbling of the jet, our images feature superluminal knots that propagate down the jet at speeds ranging from $\sim2\,c$ to $\sim13\,c$ (see Table~\ref{T3}). 

In mid-2004, the 7\,mm brightness distribution of \object{OJ287} is dominated by emission from the nearly point-like core (Fig.~\ref{maps}).
A few months later, the core region starts to expand essentially in the north-south direction.
By the end of 2005, and even more so by the end of 2007, this expansion allows us to discern two separate emission regions oriented in the north-south or northwest-southeast direction.
This relatively sudden change in the structure of the jet leads to an extremely sharp swing of JPA in mid-2004, as shown in Fig.~\ref{JPA}.
Given that this sudden jump in JPA is produced during a relatively short time period ($<1$--$2$\,yr) when the dominant emission structure is extremely compact, we propose that the event is produced by a progressive change of orientation with respect to the observer.
The innermost jet axis should have passed from one side of the line of sight (before mid-2004), to essentially along the line of sight (in mid-2004), and finally to the other side of the line of sight (after mid-2004).
This can explain the sharp, relatively short change of JPA in mid-2004, as projected on the sky in our images (Figs.~\ref{maps} and \ref{JPA}).
This interpretation is further supported by the sharp peak of core emission at about the time of the sharp JPA swing in mid-2004 (Fig.~\ref{JPA}) owing to enhanced Doppler boosting of the jet emission when the it points closer to the observer's line of sight.

Following \citet{Jorstad:2005p264}, we have estimated the variability Doppler factors ($\delta_{\rm{var}}$), viewing angles ($\theta_{\rm{var}}$), and Lorentz factors ($\Gamma_{\rm{var}}$) from the decaying flux density patterns and superluminal speeds of well-defined and bright components in our images of \object{OJ287} (Table~\ref{T4}). 
To perform estimates of these physical parameters that are as reliable as possible, we select only jet features with well-determined positions and maximum apparent speeds. 
These are associated with isolated jet features or with a leading perturbation, rather than to trailing features, which are known to display intrinsic pattern speeds much lower than those of the corresponding leading features \citep{Agudo:2001p460}.
We also require the model-fit components to be identified over more than five epochs, with peak (mean) flux densities higher than 200\,mJy (90\,mJy), and with clearly decaying flux density patterns.
The scatter in our time-dependent estimates of viewing angle $\Delta\theta_{\rm{var}}\lesssim2.7$, which we primarily attribute to small uncertainties in the determination of $\theta_{\rm{var}}$, does not allow us to verify the proposed hypothesis of the crossing of the jet from one side of the line of sight to the other.
However, the small values of the viewing angle, $\theta_{\rm{var}}\in[0.7,3.4]$, during the 1997-2010 time range favors this hypothesis, which is more likely for jets with $\theta$ close to zero.

The new brightness distribution of \object{OJ287} after the extreme JPA change in mid 2004 (see Fig.~\ref{JPA}) also shows clear wobbling of the jet from mid-2004 to mid-2011. 
Such long term wobbling, which is essentially tracked by the JPA defined by the core and the nearby, bright jet feature a, causes the jet to rotate in the plane of the sky from $\sim0^{\circ}$ at the beginning of 2005 to $\sim-50^{\circ}$ at the beginning of 2011.
Superposed on this long-term JPA variation after mid-2004, we also find clear short-term JPA variability.
Although the amplitude of these short-term fluctuations in the 2004.5-2011.5 time range ($\sim40^{\circ}$) seems to be larger than that from 1995 to 2004.5, the time scale is roughly similar in both cases ($\sim2$\,yr).

\subsection{Non-Ballistic Superluminal Motion}
\label{kin}

We use the model fits performed for every image to identify all jet features that persist over several epochs or more (Fig.~\ref{maps} and Tables~\ref{T2} and \ref{T3}). 
From these identifications, we derive the trajectories of every prominent jet feature (with average flux density $<S>\gtrsim40$\,mJy).
Among all identified components at several epochs we find both stationary and moving features, the latter showing both rectilinear (or slightly bent) and non-radial superluminal motion (see Table~\ref{T3}).
For rectilinear trajectories (with superluminal speeds in the range 2-13\,$c$), we calculate linear fits of $r$ (distance to the core) \emph{versus} $t$ (time) to estimate average apparent speeds and ejection times ($t_0$). For non-radial superluminal trajectories (showing apparent speeds up to $\sim15\,c$) we follow the second-order polynomial method of \citet{Agudo:2007p132} to fit the trajectories and determine the kinematic properties of the corresponding jet features.
Although superluminal non-radial motions have been reported before, e.g., in \object{NRAO~150} \citep{Agudo:2007p132}, this is the first time that superluminal non-radial motions $\gtrsim10\,c$ have been reported in a blazar (jet feature l, Table~\ref{T3}).

We find highly non-radial superluminal motions in the compact jet of \object{OJ287} only after mid-2004, when the projected jet axis appears to be sharply bent.
The moving jet features during the 2004.5-2011.5 time range (i.e., components {\it a}, {\it l}, {\it m}, and {\it n}, Table~\ref{T3}) follow highly bent trajectories. This demonstrates the non-ballistic motion of the jet fluid on the scales covered by our VLBA observations during this time range.
The fact that only rectilinear \emph{projected} trajectories are observed during the time range when the jet exhibits a more rectilinear shape, i.e., 1995-2004.5, does not rule out that even during this time range the fluid is non-ballistic in three dimensions.

\section{Discussion}

\subsection{Erratic versus Periodic Jet Wobbling}

Neither the short-term nor long-term variations in JPA displayed by Fig.~\ref{JPA} correspond to a simple pattern that we can relate to regular or periodic motion, at least over the time scales covered by our observations ($\sim16$\,yr). 
Instead, they seem to be related to some erratic process acting on short observed time scales $\lesssim2$\,yr with amplitudes $\lesssim40^{\circ}$ (see JPA swings during 1995, 2007, 2009, and 2011 in Fig.~\ref{JPA}).

Although based on poorer time and spatial sampling than that employed in our study, slower JPA changes than those reported here have been suggested before on larger projected scales (tens of parsecs) than those probed by our 7\,mm observations (a few parsecs); see \citet{Tateyama:2004p11950,Moor:2011p16214}, who used $3.5$\,cm observations, and \cite{Valtonen:2011p15373}, who employed VLBI observations at 2 to 6\,cm.
VLBA observations at similar radio wavelengths (2\,cm) compiled by the MOJAVE team\footnote{\tt www.physics.purdue.edu/astro/MOJAVE/animated/ 0851$+$202.i.mpg} \citep{Lister:2009p5316} over more than 10 years also supports the results of \citet{Tateyama:2004p11950,Moor:2011p16214,Valtonen:2011p15373}. 
However, the periodic precession model proposed by \citet{Tateyama:2004p11950} predicted a jet structural position angle $\sim-100^{\circ}$ in 2007, whereas we measure a drastically different inner JPA of $\sim0^{\circ}$.
Hence, the jet structure shown in Fig.~\ref{maps} does not follow Tateyama \& Kingham's 12\,yr periodic precession model.
The JPA observed by \citet{Moor:2011p16214} is also $\gtrsim100^{\circ}$ off our 7\,mm measurements on similar dates.
Moreover, the JPA evolution on the scales observed through our observations (Fig.~\ref{JPA}) conflicts with the JPA evolution model presented by \citet{Valtonen:2011p15373}.
This model predicts a JPA value in the range $-110^{\circ}$ to $-150^{\circ}$ at the beginning of 2009, whereas we measure $\sim-10^{\circ}$.

This apparent inconsistency could be reflecting a (perhaps periodic) long-term JPA variation in the jet on larger scales than those probed by our 7\,mm VLBA observations, together with more erratic behavior modulating the long-term JPA variations in the innermost jet regions probed by our millimeter monitoring.
It thus appears that a single phenomenon cannot explain both the JPA variations observed on the scales of tens of parsecs, and the much shorter term ($\sim2$\,yr) JPA variations reported in this work. 
This implies that processes other than those producing the long-term JPA variations at large jet scales need to take place near the jet origin to modulate the jet direction on shorter temporal and spatial scales.

\subsection{Possible Origin of Observed Jet Wobbling}

Different explanations have been proposed for wobbling in blazar jets.
Orbital motion of compact objects in a binary black hole system has been suggested as a possibility to explain bends in AGN jets \citep{Begelman:1980p16092}.
Alternatively, accretion disk precession could arise from either a companion super-massive black hole or another massive
object inducing torques in the accretion disk of the primary \citep[e.g.,][]{2003ApJ...584..135L}, or by the Lense-Thirring effect \citep{Bardeen:1975p16093} in the presence of a Kerr black hole with its rotational axis misaligned with regard to that of the accretion disk \cite[e.g.][]{Caproni:2004p13810}.
However, these processes cannot, by themselves, explain the short time scales ($\sim2$\,yr) and erratic nature of the JPA changes reported here.

Pressure gradients in the ambient external medium are also not, by themselves, a feasible explanation for the short-term JPA variations reported here. 
They might, however, explain the smoothly bent jet structure observed on scales of hundreds of kiloparsecs \citep{Marscher:2011p15116}. 

A geometrical model based on non-ballistic motions of blobs and misidentification of components has also been
proposed as a possible explanation of JPA variations in general \citep{Gong:2008p16025}. 
However, we can exclude misidentification in our observations of \object{OJ287} owing to the good time sampling of our monitoring. 

Other kinds of more erratic disk or jet instabilities \citep[e.g., similar to those produced by turbulent accretion;][]{McKinney:2009p12064} have not been extensively explored previously as an explanation of jet wobbling in blazar jets.
Such instabilities could explain the apparently erratic JPA variations that modulate the behavior of the innermost jet of \object{OJ287}. 
We discuss this possibility in the following subsections.

\subsection{Jet Wobbling Produced by Variations of Flow Injection}
\label{inj}

Since the fast and erratic JPA modulations that we report in this paper are observed in the innermost jet regions, we propose that time variations occur in the injection of the jet flow at the origin of the jet, perhaps resulting from changes in the particle density or magnetic field configuration caused by turbulence or irregularities in the accretion process. 
Such variations, combined with the axial rotation of the jet material in the presence of significant toroidal magnetic field components, can produce non periodic JPA variations \citep{McKinney:2009p12064}. 
If transverse gradients in pressure and velocity at the jet injection persist for extended time intervals, they should generate bent jet structures \citep[e.g.,][]{Aloy:2003p350}, that---amplified by projection effects from small mean jet viewing angles---may produce sharp apparent bends such as those observed in \object{OJ287} (see Fig.~\ref{polmaps}).

\subsection{Large Apparent JPA variations and Sharp Jet Bends}

The small viewing angle of the jet axis in some sources allows us to see the jet from inside its cone, in which case asymmetric changes in the jet can cause it to appear to cross the line of sight.
The sharp jump in JPA in mid-2004 from $\rm{JPA}\approx-150^{\circ}$ to $\rm{JPA}\approx0^{\circ}$ seen in our 7\,mm sequence of images supports this scenario (Section \ref{OJ287wobbling}).
This requires only minor changes of the jet structure provided that the mean viewing angle of the jet is close to $0^{\circ}$, as we show for \object{OJ287} (see Table~\ref{T4}). 

Projection effects from very small viewing angles can also create the illusion of very sharp bends from slightly bent jet structures, as is the case of the main bend in the jet of \object{OJ287} from north-northeast to southeast (see Fig.~\ref{polmaps}).
This bend could result from jet instabilities \citep{Hardee:2011p16037,Perucho:2011p16709} causing the jet flow to curve.
Such instabilities can be triggered by variations at the jet injection as those mentioned in Section~\ref{inj}.
Moreover, it is reasonable to expect pressure gradients in the external medium to play an important role in re-directing the northeasterly flow toward the previous JPA of $\sim -110^\circ$ at the edge of the channel carved out by the jet over thousands of years. 
Such bending would be accomplished via multiple oblique shocks at the outer boundary of the jet. 
This in turn would be accompanied by compression of the magnetic field and acceleration of relativistic electrons, which would (1) enhance the synchrotron emission near the outer edge and 
(2) align the magnetic field near the outer edge so that it follows the curvature of the jet. 
The polarization map of 2011-04-21 (and possibly 2011-06-12; see Fig.~\ref{polmaps}) conform with these expectations, although the angular resolution in the north-south direction is inadequate to confirm the predicted edge-brightening.

\subsection{Relation to Long Scales}

If inhomogeneous jet injection occurs randomly, i.e., with no systematic temporal pattern, its effect on the jet flow in terms of instabilities would decrease with distance and within the jet cross section. 
Indeed, the mean long-term position angle of the compact jet found by \citet{Tateyama:2004p11950} is $\sim -107^\circ$, very similar to the value of $-109^\circ$ of the kiloparsec-scale jet within $8''$ of the nucleus \citep{Marscher:2011p15116}. 
This implies that only short wavelengths are excited, perhaps in the form of high-order body modes. 
Such short wavelengths grow rapidly in amplitude but are inefficient in jet mixing and disruption in the non-linear phase, in contrast with instabilities that generate large deformations on the jet surface, i.e., first-order body and surface modes \citep{Xu:2000p16069,Perucho:2004p16052,Perucho:2004p16053,Perucho:2005p336,Perucho:2007p15976,Perucho:2010p16057,Hardee:2011p15692}. 
Thus, the small-scale nature of the observed changes in our 7\,mm VLBA data do not imply disruption of the jet on larger scales despite the high apparent amplitudes, which are accentuated by projection effects.

\section{Conclusions}

Our 7\,mm VLBA study has revealed a sharp jet-position-angle swing by $>100^{\circ}$ during [2004,2006], as viewed in the plane of the sky, that we interpret as the crossing of the jet from one side of the line of sight to the other during a softer and longer term swing of the inner jet.
Modulating such long term swing, our images also show that the innermost $\sim0.4$\,mas region of the 7\,mm jet in \object{OJ287} wobbles following an erratic pattern with amplitudes up to $\sim40^{\circ}$ on time scales of $\sim2$\,yr.
The latter phenomenon, reported here for the first time, rules out scenarios such as the putative orbit of the supermassive black hole responsible for the production of the jet in a binary system, accretion disk precession induced by either a companion compact object or the Lense-Thirring effect, and jet interaction with the external medium as possible origins of the phenomenon. 
This is based on the long time scales related to these scenarios compared with the $\sim2$\,yr time scale and to the apparently erratic nature of the observed wobbling in \object{OJ287}. 
The dense time sampling of our observations also allows us to rule out misidentification of jet features across epochs as a possible explanation.
This leads us to propose a new possibility to explain the non-periodic jet wobbling reported here through asymmetric fluctuations in injection of plasma flow into the jet coupled with instabilities. 
This could be related to variations in the accretion process and in the disk-jet connection. 

Both our observations and those from other VLBI programs at centimeter wavelengths \citep{Moor:2011p16214,Valtonen:2011p15373} rule out the $\sim12$\,yr periodic precession model proposed by \citet{Tateyama:2004p11950}, which predicted different evolution of the JPA than observed.
In contrast, the 120\,yr periodicity of the \citet{Valtonen:2011p15373} precession model seems able to reproduce the JPA evolution observed at centimeter wavelengths, although the 12\,yr modulation that their model imparts on the predicted JPA changes does not clearly follow their observed JPA variations during the [2005,2011] time range.
Such 12\,yr modulation might be better reproduced by the $3.5$\,cm JPA observations by \citet{Moor:2011p16214}, but more data is still needed to clarify the consistency between new Valtonen's JPA evolution-model and the centimeter VLBI jet behavior. 

Our interpretation of the apparent mismatch between the JPA evolution reported in this paper (i.e., on scales of a few parsecs) and those observed at centimeter wavelengths (tens of parsecs) leaves open the possibility that different instability modes operate on these two spatial scales.
This allows for the JPA on the inner scales to behave in an erratic way, while on the larger scales it can be driven by instabilities coupled to modes with periodic behavior.
Periodic models based on binary black hole systems are thus not ruled out by our observations, unless future centimeter VLBI observations of the jet in \object{OJ287} show more erratic JPA variations similar to those seen at millimeter wavelengths.

The detailed study of the evolution of the JPA of \object{OJ287} presented here, and the main conclusions drawn above, were only possible with the ultra-high resolution provided by the VLBA at 7\,mm and long-term monitoring of \object{OJ287} with dense time sampling.
Studies similar to that presented here can be carried out for a number of blazars that have been monitored intensively with millimeter VLBI during the past decade. 
Performing these studies will be crucial to evaluate how common erratic wobbling is in the jets of AGN, and to draw general conclusions about the possible relation of this phenomenon to the accretion process.

\begin{acknowledgements}
This research has made use of the MOJAVE database maintained by the MOJAVE team \citep{Lister:2009p5316}.
The VLBA is an instrument of the National Radio Astronomy Observatory, a facility of the National Science Foundation operated under cooperative agreement by Associated Universities, Inc. 
The research at the IAA-CSIC is supported in part by the Ministerio de Ciencia e Innovaci\'{o}n of Spain, and by the regional government of Andaluc\'{i}a through grants AYA2010-14844 and P09-FQM-4784, respectively. 
The research at Boston University was funded by US National Science Foundation grant AST-0907893, NASA grants NNX08AJ64G, NNX08AU02G, NNX08AV61G, and NNX08AV65G, and NRAO award GSSP07-0009. 
\end{acknowledgements}


\input{./Table1}

\input{./Table2}

\input{./Table3}

\input{./Table4}

\end{document}

%% file: Table1.tex
\begin{deluxetable}{lccccccc}
\tablecolumns{8}
\tabletypesize{\scriptsize}
\tablewidth{0pt} 
\tablecaption{\label{T1} Observing log and image information. (Truncated table).}
\tablehead{\colhead{Epoch} & \colhead{$\rm{t}_{\rm{int}}$} & \colhead{Bit} & 
                  \colhead{$\Delta\nu_{\rm{obs}}$} & \colhead{$S_{\rm{int}}$} &   
                  \colhead{$S_{\rm{peak}}$} & \colhead{Noise} & \colhead{Ref.}\\
                  \colhead{} & \colhead{min} &  \colhead{samp} & \colhead{MHz} & 
                  \colhead{Jy}  &  \colhead{Jy/beam} & \colhead{mJy/beam} & \colhead{}\\
                  \colhead{(1)} &  \colhead{(2)} &  \colhead{(3)} &  \colhead{(4)} &  
                  \colhead{(5)} &  \colhead{(6)} & \colhead{(7)} &  \colhead{(8)}}
\startdata
 1995--01--05   & 68  &  1   &   32  &  1.01 &  0.87 & 2.5 &  1    \\ 
 1995--04--18   & 25  &  1   &   32  &  1.29 &  0.93 & 3.1 &  1    \\ 
 1995--06--04   & 34  &  1   &   32  &  1.15 &  0.93 & 1.8 &  1    \\ 
 1995--10--15   & 14  &  1   &   32  &  2.57 &  1.93 & 4.4 &  1    \\ 
 1996--05--03   & 30  &  1   &   64  &  1.18 &  0.90 & 1.6 &  1    \\ 
 1996--11--24   & 35  &  1   &   64  &  1.19 &  0.85 & 1.3 &  1    \\ 
 1997--07--30   & 35  &  1   &   64  &  1.45 &  1.09 & 0.8 &  1    \\ 
 1997--11--10   & 30  &  1   &   64  &  1.48 &  1.06 & 0.6 &  2    \\ 
 1997--12--11   & 32  &  1   &   64  &  1.26 &  0.81 & 0.6 &  2    \\ 
 1998--01--11   & 32  &  1   &   64  &  1.58 &  1.13 & 1.3 &  2    \\ 
 1998--02--07   & 32  &  1   &   64  &  1.83 &  1.38 & 0.7 &  2    \\ 
\enddata
\tablecomments{Columns are as follows: 
                         (1) observing epoch, 
                         (2) total integration time, 
                         (3) number of bits used for signal digitalization sampling, 
                         (4) observing frequency bandwidth, 
                         (5) total integrated flux density, 
                         (6) peak flux density, 
                         (7) noise level of the resulting image,
                         (8) reference where data was published.}
\tablerefs{(1) \cite{Jorstad:2001p5655};                     
           (2) \cite{Gomez:2001p201};                            
           (3) \cite{Agudo:2006p331};                           
           (4) \cite{Bach:2006p354};                           
           (5) \cite{Gomez:2008p30};                            
           (6) \cite{Gomez:2011p15108};                            
           (7) \cite{Piner:2006p12065};                            
           (8) data from VLBA archive;                            
           (9) \cite{Agudo:2007p132};                            
           (10) \cite{Darcangelo:2009p6953};                            
           (11) \cite{Agudo:2010p12401};                            
           (12) \cite{Agudo:2011p14707};
           (13) this paper.}
\end{deluxetable}

%% file: Table2.tex
\begin{deluxetable}{lcccc}
\tablecolumns{5}
\tabletypesize{\scriptsize}
\tablewidth{0pt}
\tablecaption{\label{T2} Model fit parameters. (Truncated table).}
\tablehead{\colhead{Comp} & \colhead{$S$} & \colhead{$r$} & \colhead{$\theta$} & \colhead{FWHM}\\
           \colhead{ } & \colhead{mJy} & \colhead{mas} & \colhead{$^{\circ}$} & \colhead{mas}\\
           \colhead{(1)} & \colhead{(2)} & \colhead{(3)} & \colhead{(4)} & \colhead{(5)}}
\startdata
\hline\noalign{\smallskip}
\multicolumn{5}{c}{1995-01-05}\\
\hline\noalign{\smallskip}
C & $ 902\pm 60$ & $0.00\pm0.00$ & $   0\pm   0$ & $0.022\pm0.001$ \\
A & $  96\pm  7$ & $0.23\pm0.03$ & $ -83\pm   7$ & $0.176\pm0.009$ \\
B & $  55\pm  4$ & $0.62\pm0.03$ & $ -88\pm   2$ & $0.243\pm0.012$ \\
\hline\noalign{\smallskip}
\multicolumn{5}{c}{1995-04-18}\\
\hline\noalign{\smallskip}
C & $1034\pm 69$ & $0.00\pm0.00$ & $   0\pm   0$ & $0.061\pm0.001$ \\
A & $ 210\pm 15$ & $0.24\pm0.03$ & $ -95\pm   7$ & $0.130\pm0.006$ \\
B & $  54\pm  3$ & $0.78\pm0.03$ & $ -94\pm   2$ & $0.259\pm0.013$ \\
D & $  63\pm  4$ & $1.09\pm0.03$ & $ -90\pm   1$ & $0.186\pm0.009$ \\
E & $  91\pm 10$ & $1.40\pm0.25$ & $ -92\pm  10$ & $0.510\pm0.051$ \\
\hline\noalign{\smallskip}
\multicolumn{5}{c}{1995-06-04}\\
\hline\noalign{\smallskip}
C & $1034\pm 69$ & $0.00\pm0.00$ & $   0\pm   0$ & $0.049\pm0.001$ \\
A & $  41\pm  4$ & $0.30\pm0.01$ & $-118\pm   1$ & $0.000\pm0.001$ \\
D & $ 142\pm 17$ & $1.11\pm0.29$ & $ -91\pm  14$ & $0.570\pm0.057$ \\
\enddata
\tablecomments{Columns are as follows:
                         (1) jet feature,
                         (2) flux density,
                         (3) distance to the core,
                         (4) position angle,
                         (5) FWHM.}
\end{deluxetable}

%% file: Table3.tex
\begin{deluxetable}{lccccccccc}
\tablecolumns{10}
\tabletypesize{\scriptsize}
\tablewidth{0pt} 
\tablecaption{\label{T3} Properties of model fit components.}
\tablehead{\colhead{Comp} & \colhead{$<S>$} & \colhead{$<\theta>$} & \colhead{$<r>$} & \colhead{$<v_{\rm{app}}>$} & \colhead{$<v_{\rm{app}}>$} & \colhead{$t_{0}$} & \colhead{$<v_{\rm{app}}^{\rm{non radial}}>$} & \colhead{$<v_{\rm{app}}^{\rm{non radial}}>$} & \colhead{$<\dot\Theta>$} \\
           \colhead{ } & \colhead{mJy} & \colhead{$^{\circ}$} & \colhead{mas} & \colhead{mas/yr} & \colhead{$c$} & \colhead{yr} & \colhead{mas/yr} & \colhead{$c$} & \colhead{$^{\circ}/\rm{yr}$} \\   
           \colhead{(1)} & \colhead{(2)} & \colhead{(3)} & \colhead{(4)} & \colhead{(5)} & \colhead{(6)} & \colhead{(7)} & \colhead{(8)} & \colhead{(9)} & \colhead{(10)}}
\startdata
\hline\noalign{\smallskip}
\multicolumn{10}{c}{Quasi-stationary or slowly moving features}\\
\hline\noalign{\smallskip}
  A\tablenotemark{a}     &  $294\pm182$ & $-113\pm 8$ & $0.28\pm0.06$ &      \nodata      &     \nodata        &    \nodata  &             \nodata                      &            \nodata                  &        \nodata         \\
  F     &  $ 47\pm 25$ & $ -98\pm 6$ & $1.38\pm0.11$ &      \nodata      &     \nodata        &    \nodata  &            \nodata                       &            \nodata                  &        \nodata         \\
  G\tablenotemark{b}     &  $ 77\pm 41$ & $-110\pm 2$ & $1.02\pm0.06$ &      \nodata      &     \nodata        &    \nodata  &            \nodata                       &            \nodata                  &        \nodata         \\
  L\tablenotemark{c}     &  $174\pm148$ & $-102\pm15$ & $0.15\pm0.04$ &      \nodata      &     \nodata        &    \nodata  &            \nodata                       &            \nodata                  &        \nodata         \\
  e     &  $158\pm111$ & $ -89\pm19$ & $0.26\pm0.07$ &      \nodata      &     \nodata        &    \nodata  &            \nodata                       &            \nodata                  &        \nodata         \\
  f     &  $400\pm228$ & $ -70\pm16$ & $0.19\pm0.03$ &      \nodata      &     \nodata        &    \nodata  &            \nodata                       &            \nodata                  &        \nodata         \\
  g     &  $ 89\pm 36$ & $-127\pm 9$ & $0.74\pm0.03$ &      \nodata      &     \nodata        &    \nodata  &            \nodata                       &            \nodata                  &        \nodata         \\
  k     &  $200\pm138$ & $ -57\pm12$ & $0.18\pm0.02$ &      \nodata      &     \nodata        &    \nodata  &            \nodata                       &            \nodata                  &        \nodata         \\
\hline\noalign{\smallskip}
\multicolumn{10}{c}{Superluminal rectilinear features}\\
\hline\noalign{\smallskip}
  H\tablenotemark{d}     &  $ 93\pm 63$ & $-116\pm 6$ & $0.43\pm0.08$ &  $0.32\pm0.08$ &   $6.0\pm1.5$   & $1997.71\pm0.25$ &            \nodata                      &            \nodata                  &        \nodata        \\ 
  J\tablenotemark{e}     &  $119\pm 99$ & $-111\pm 3$ & $0.57\pm0.16$ &  $0.40\pm0.07$ &   $7.7\pm1.3$   & $1998.02\pm0.24$ &            \nodata                       &            \nodata                  &        \nodata        \\ 
  K\tablenotemark{f}     &  $ 37\pm 24$ & $-113\pm 8$ & $0.79\pm0.34$ &  $0.41\pm0.04$ &   $7.7\pm0.7$   & $1999.22\pm0.18$ &            \nodata                       &            \nodata                  &        \nodata        \\ 
  O\tablenotemark{g}     &  $243\pm135$ & $-116\pm 4$ & $1.19\pm0.26$ &  $0.50\pm0.01$ &   $9.5\pm0.1$   & $2000.16\pm0.03$ &            \nodata                       &            \nodata                  &        \nodata        \\ 
  N     &  $288\pm152$ & $-122\pm 2$ & $0.58\pm0.09$ &  $0.39\pm0.14$ &   $7.5\pm2.6$   & $2000.76\pm0.33$ &            \nodata                       &            \nodata                  &        \nodata        \\ 
  R     &  $ 91\pm 16$ & $-123\pm 1$ & $0.85\pm0.18$ &  $0.68\pm0.06$ &  $13.0\pm1.2$   & $2001.92\pm0.12$ &            \nodata                       &            \nodata                  &        \nodata        \\ 
  T     &  $ 97\pm 46$ & $-124\pm 6$ & $0.97\pm0.30$ &  $0.40\pm0.08$ &   $7.6\pm1.5$   & $2003.22\pm0.38$ &            \nodata                       &            \nodata                  &        \nodata        \\ 
  V     &  $221\pm314$ & $-130\pm10$ & $0.51\pm0.44$ &  $0.54\pm0.05$ &  $10.2\pm0.9$   & $2004.30\pm0.10$ &            \nodata                       &            \nodata                  &        \nodata        \\ 
  b     &  $ 72\pm 36$ & $-120\pm 3$ & $1.17\pm0.13$ &  $0.46\pm0.15$ &   $8.8\pm2.8$   & $2006.34\pm0.48$ &            \nodata                       &            \nodata                  &        \nodata        \\ 
  d     &  $ 71\pm 46$ & $-126\pm14$ & $0.60\pm0.24$ &  $0.24\pm0.01$ &   $4.5\pm0.2$   & $2004.77\pm0.13$ &            \nodata                       &            \nodata                  &        \nodata        \\ 
  h     &  $120\pm 50$ &  $-98\pm22$ & $0.25\pm0.09$ &  $0.30\pm0.02$ &   $5.7\pm0.4$   & $2006.78\pm0.06$ &            \nodata                       &            \nodata                  &        \nodata        \\ 
  j     &  $157\pm 49$ & $-126\pm 7$ & $0.81\pm0.39$ &  $0.35\pm0.02$ &   $6.7\pm0.4$   & $2007.85\pm0.12$ &            \nodata                       &            \nodata                  &        \nodata        \\ 
\hline\noalign{\smallskip}
\multicolumn{10}{c}{Superluminal bent features}\\
\hline\noalign{\smallskip}
  a     & $2293\pm1631$  & $-22\pm23$  &  $0.17\pm0.06$ &  $0.07\pm0.01$\tablenotemark{h}  &$1.4\pm0.1$\tablenotemark{h}     &    \nodata  &        $0.07\pm0.01$\tablenotemark{h}             &             $1.4\pm0.1$\tablenotemark{h}           &   $-19\pm1$\tablenotemark{h}  \\
  l     & $ 444\pm 404$  & $-62\pm21$  &  $0.33\pm0.07$ &    $0.94\pm0.21$    & $17.9\pm4.0$       &    \nodata  &          $0.78\pm0.29$               &               $14.9\pm5.5$            &   $-145\pm40$    \\
  m     & $ 463\pm 284$  & $-75\pm 9$  &  $0.35\pm0.04$ &    $0.62\pm0.14$    & $11.9\pm2.6$       &    \nodata  &          $0.52\pm0.18$               &               $10.0\pm3.5$            &    $-87\pm21$    \\
  n     & $ 531\pm 149$  & $-67\pm13$  &  $0.40\pm0.13$ &    $0.54\pm0.02$    & $10.3\pm0.5$       &    \nodata  &          $0.22\pm0.09$               &                $4.3\pm1.7$            &     $-40\pm4$    \\
\enddata
\tablecomments{Columns are as follows: 
                         (1) jet feature, 
                         (2) average flux density, 
                         (3) average position angle, 
                         (4) average distance to the core, 
                         (5) average proper motion in mas/yr, 
                         (6) average proper motion in units of $c$, 
                         (7) estimated ejection time if any,
                         (8) average proper motion in non radial direction in mas/yr,
                         (9) average proper motion in non radial direction in units of $c$,
                         (10) average angular speed.}
\tablenotetext{a}{This jet feature corresponds to stationary component A2 as identified by \citet{Jorstad:2005p264}, and C1 as identified by \citet{Jorstad:2001p5655}.}
\tablenotetext{b}{This jet feature corresponds to stationary component A3 as identified by \citet{Jorstad:2005p264}.}
\tablenotetext{c}{This jet feature corresponds to stationary component A1 as identified by \citet{Jorstad:2005p264}.}
\tablenotetext{d}{This feature may correspond in some epochs to superluminal knot B4 identified by \citet{Jorstad:2005p264}. We prefer our identification owing to our better time sampling during 1998 though.}
\tablenotetext{e}{This feature corresponds to superluminal knot B5 identified by \citet{Jorstad:2005p264}, although they reported faster superluminal speed ($12\pm2\,c$) perhaps owing to their better time sampling during 1999, hence producing better kinematic estimates.
}
\tablenotetext{f}{This feature may correspond to superluminal knot B6 identified by \citet{Jorstad:2005p264}. We prefer our identification owing to our better time sampling during 2001 though.}
\tablenotetext{g}{This feature corresponds to superluminal knot B7 reported by \citet{Jorstad:2005p264}.}
\tablenotetext{h}{Computed only for the time range from 2008-10-22 to 2010-10-24, when the bent trajectory of component a is better defined and reaches superluminal non-radial speeds.}
\end{deluxetable}

%% file: Table4.tex
\begin{deluxetable}{lcccc}
\tablecolumns{5}
\tabletypesize{\scriptsize}
\tablewidth{0pt} 
\tablecaption{\label{T4} Physical parameters of well defined superluminal components}
\tablehead{\colhead{Comp} & \colhead{$t_{0}$} & \colhead{$\delta_{\rm{var}}$} &  \colhead{$\theta_{\rm{var}}$} &  \colhead{$\Gamma_{\rm{var}}$}  \\
           \colhead{ } & \colhead{yr} & \colhead{ } & \colhead{deg.} & \colhead{ } \\   
           \colhead{(1)} & \colhead{(2)} & \colhead{(3)} & \colhead{(4)} & \colhead{(5)}}
\startdata
H                  &  $1997.71\pm0.25$ & 15.5 & 2.7 &  8.7\\
J\tablenotemark{a} &  $1998.61\pm0.09$ & 16.1 & 3.4 & 12.3\\   
O                  &  $2000.16\pm0.03$ & 33.3 & 0.9 & 18.0\\
V                  &  $2004.30\pm0.10$ & 22.6 & 1.9 & 13.6\\
h                  &  $2006.78\pm0.06$ & 25.2 & 1.0 & 13.3\\
n                  &  $>2010.64$ & 36.2 & 0.7 & 19.1\\
\enddata
\tablenotetext{a}{We show the parameters computed by \citet{Jorstad:2005p264}, who made a better determination of the kinematics of this jet feature (see Table~\ref{T3}).}
\end{deluxetable}